\documentclass[aps,prd,reprint,nofootinbib,amsmath,amssymb,floatfix]{revtex4-2}
\usepackage{fixes}
\usepackage[table]{xcolor}            
\definecolor{PRDblue}{HTML}{2E3092}
\usepackage[colorlinks=true, allcolors=PRDblue]{hyperref} 
\usepackage{pgfplots}           
\pgfplotsset{compat=1.18}       
\usetikzlibrary{arrows.meta,decorations.markings,calc}
\usetikzlibrary{external}       
\usepackage{tikzscale}          
\graphicspath{{figures/}}       
\usepackage{siunitx}            
\usepackage{graphicx}	        
\usepackage{physics}            
\usepackage{cleveref}           
\usepackage{isomath}            %
\usepackage{dutchcal}           
\usepackage{silence}
\providecommand{\ii}{\mathrm{i}}
\providecommand{\ee}{\mathrm{e}}
\providecommand{\vc}{\vb*}
\providecommand{\uv}{\vu*}
\renewcommand{\Re}{\real}
\renewcommand{\Im}{\imaginary}
\newcommand{\rc}{\mathcal{r}}
\newcommand{\vcrc}{\vc{\mathcal{r}}}
\newcommand{\uvrc}{\uv{\mathcal{r}}}

\begin{document}
\title{On the origin and properties of dipolar recoil force and torque}
\author{Sebastian \surname{Golat}}   
\thanks{These two authors contributed equally}
\author{Nathaniel \surname{Levy}}       
\thanks{These two authors contributed equally}
\author{Francisco J. \surname{Rodr\'iguez-Fortu\~no}} \email{francisco.rodriguez\_fortuno@kcl.ac.uk}
\affiliation{
Department of Physics and London Centre for Nanotechnology, 
King's College London, Strand, London WC2R 2LS, UK}

\date{\today}
\begin{abstract}
The recoil optical force and torque acting on an electromagnetic dipole are typically derived by computing the imbalance in radiated linear and angular electromagnetic momentum coming from the source, using Maxwell stress tensor integration. This quantifies the recoil's outcome without revealing its physical origin or the underlying forces that produce it. The recoil force and torque exist even in the absence of external illumination, such that an isolated dipole emitter can experience them. In contrast to the other terms, the recoil terms are odd under time reversal. To clarify their nature and properties, we re-derive the recoil force from first principles using the total Lorentz force on a system of charges that form a simultaneous electric and magnetic dipole. The results agree with the standard momentum-based derivation and reveal their ultimate origin from retardation effects---arising from the finite speed of light---in the mutual interactions between charges. This result provides fundamental insight into the recoil mechanism, offering a clearer conceptual foundation for future theoretical and experimental studies of light-induced forces.
\end{abstract}

\maketitle%
\section{Introduction}%
{Light can exert mechanical action on matter, enabling optical trapping and manipulation \cite{Dholakia2011Jun,Grier2003N,Dienerowitz2008}, laser cooling and atom manipulation \cite{Ashkin2000IEEE}, opto-mechanical control of micro- and nano-objects \cite{Marago2013NN,Aspelmeyer2014RMP,Millen2020RPP,Gonzalez-Ballestero2021S}, and spin- and orbital-angular-momentum–driven forces and torques in structured electromagnetic fields \cite{Sukhov2017RPP,Yang2021AP}, among other applications}. When a material system is small compared to the wavelength of light, light-matter interactions can often be described via a dipolar model. Thus the expressions for forces and torques acting on a dipolar particle form the foundation of many optical force applications.

The time-averaged optical force $ \langle \vc{F} \rangle$ acting on a time-harmonic dipolar particle is \cite{Nieto-Vesperinas2010, Chaumet2009, Albaladejo2009, Gao2017, Toftul2024Oct}
\begin{equation}\label{eq:dipoleforce}
\!\!\!\!\langle{\vc{F}}\rangle\!=\!\underbrace{\tfrac{1}{2}\Re\big[(\grad\!\otimes\!\vc{E})\!\vdot\!\vc{p}^\ast\!+\!(\grad\!\otimes\!\vc{B})\!\vdot\!\vc{m}^\ast\big]}_\text{interaction $\langle{\vc{F}_\text{int}}\rangle$}\!
    \underbrace{-\tfrac{k^4\eta}{12\pi}\Re(\vc{p}^\ast\!\!\cp\!\vc{m})}_\text{recoil $\langle{\vc{F}_\text{rec}}\rangle$},\!\!
\end{equation}%
and the torque $ \langle \vc{\varGamma} \rangle$ on the same particle is \cite{wei2022optical,Toftul2024Oct,Chaumet2009, Canaguier-Durand2013PRA, Bliokh2014a, Bliokh2015, Nieto-Vesperinas2015OL}
\begin{equation}\label{eq:theoreticaltorque}
\begin{split}
    \!\!\!\!\!\langle \vc{\varGamma} \rangle \!=& \underbrace{\tfrac{1}{2} \Re\big(\vc{p}^* \!\!\cp \vc{E} +  \vc{m}^* \!\!\cp \vc{B}\big)}_\text{interaction $\langle{\vc{\varGamma}_\text{int}}\rangle$}\\
    &
    \hphantom{\!\underbrace{\tfrac{1}{2} \Re\big(\vc{p}^* \!\!\cp \vc{E} \big)+\vc{m}}_\text{interaction}}
    \underbrace{ -\tfrac{k^3}{12 \pi\varepsilon } \Im\qty(\vc{p}^* \!\!\cp \vc{p} + \tfrac{1}{c^2} \vc{m}^* \!\!\cp \vc{m} )}_\text{recoil $\langle{\vc{\varGamma}_\text{rec}}\rangle$} ,
\end{split}
\end{equation}
with $\vc{p}$ and $\vc{m}$ being the electric and magnetic dipole moments existing at the same point in space, and $\vc{E}$ and $\vc{B}$ are the external illumination fields evaluated at that point, all considered as time-harmonic phasors. Both force and torque can be divided into two distinct parts. The first group is known as the interaction (extinction) terms, in which the dipole moments interact with the gradients of the applied field or the fields themselves. This term includes both pressure and gradient forces, and constitute the foundation of optical tweezers \cite{Dholakia2011Jun,Grier2003N,Dienerowitz2008, Ashkin_book, Ashkin1986, Ashkin2000IEEE, Ashkin1987, Ashkin1987_II}. The second terms are the recoil (scattering) force and torque, which involve only the dipole moments and not the applied fields. {These terms are often (but not always) smaller in magnitude. They are crucial, for instance, in optical tractor beams \cite{Brzobohaty2013NP}, in lateral spin-induced forces \cite{Antognozzi2016Aug}, and in chiral enantiomer separation schemes based on evanescent fields \cite{Hayat2015,Golat2024Apr}.} The recoil terms are peculiar because they exist \emph{even in the absence} of external illumination, and correspond to the force and/or torque that the magnetic and electric dipoles exert on themselves and each other. Thus, an isolated system (such as a spacecraft) could propel and rotate itself by inducing electric and magnetic dipoles within it. A naive interpretation would suggest that this breaks the law of action and reaction and the conservation of momentum; however, it is well known \cite{Jackson1998, Griffiths2012} that the electromagnetic Lorentz force between charged particles indeed breaks the law of action and reaction, while conserving \emph{total} momentum, due to the electromagnetic field momentum that is radiated. Thanks to the recoil terms, an isolated system can in principle propel and rotate itself radiating electromagnetic fields directionally and with circularly polarised spinning fields, as electric and magnetic dipole combinations do, essentially using light's linear and angular momenta as propulsion.

Indeed, the derivation of the recoil terms \cite{Nieto-Vesperinas2010} was first accomplished through the integration of the Maxwell stress tensor, which involves the calculation of the net linear and angular electromagnetic momentum flowing out of a volume per unit of time via a flux integral of the momentum flux density on any closed surface surrounding the dipoles. {This quantifies the force and torque, but does not explain the underlying forces that produce them.} It is also interesting to note the peculiar fact that the recoil force term has a different behaviour under time reversal compared to the interaction term. {The interaction force term (including pressure and gradient forces) is time-reversal even, while the recoil force term is time-reversal odd.} This {unintuitive fact} has been overlooked and is surprising, considering that the Lorentz force expression, {which is ultimately the source of all light-matter mechanical interactions,} is time-reversible. In fact, a similar mismatch occurs for the absorbed power \cite{Toftul2024Oct,Nieto-Vesperinas2010}:
\begin{equation}\label{eq:dipolepower}
\!\!\!\langle{P_\text{abs}}\rangle\!=\!\underbrace{\tfrac{\omega}{2}\Im\qty(\vc{E}^\ast\!\!\vdot\vc{p}+\!\vc{B}^\ast\!\!\vdot\vc{m})}_\text{extinction $\langle{P_\text{ext}}\rangle$}-\underbrace{\tfrac{\omega k^3}{12\pi}(\tfrac{1}{\varepsilon}\abs{\vc{p}}^2\!+\!\mu\abs{\vc{m}}^2)}_\text{scattering $\langle{P_\text{sca}}\rangle$}\,,\!\!
\end{equation}%
where $\langle{P_\text{ext}}\rangle$ is time reversal odd and $\langle{P_\text{sca}}\rangle$ is even.

In this work, we aim to clarify the origin and properties of the dipolar recoil force by investigating how it arises from the foundational electromagnetic Lorentz forces between the charges and/or currents that constitute the dipole.
Our calculation matches the recoil term of the known \cref{eq:dipoleforce}, as expected given that the Lorentz force is ultimately at the root of the Maxwell stress tensor method used to calculate them. A careful calculation of the Lorentz force exerted by each particle on every other requires considering the retarded position of each source charge as seen from the charge experiencing the force. We reveal this retardation effect to be the origin of the recoil force. This resembles the origin of the self-force (also called radiation reaction or Abraham–Lorentz force) of a charged particle on itself, which is enabled by retardation effects \cite{Griffiths2012, Jackson1998,Boyer1972Dec,Griffiths1983Dec}. In addition, the fact that the force depends only on the retarded positions and not also on the advanced ones explains the break in the time-reversal symmetry.

\section{Symmetry of the force terms under time-reversal}\label{sec:symmetries}
It is quite interesting to look at the symmetry properties of 
\cref{eq:dipoleforce,eq:theoreticaltorque,eq:dipolepower}, in particular the time reversal. These expressions contain field and dipole phasors. The relationship between a time harmonic physical quantity $\vc{\mathcal{A}}(t)$ and its phasor representation $\vc{A}$ is 
$\vc{\mathcal{A}}(t)=\Re(\vc{A}\ee^{-\ii\omega t})$. One can easily confirm that if the quantity is either even or odd under time reversal:
\begin{equation}
    \vc{\mathcal{A}}_\text{even/odd}(-t)=\pm \vc{\mathcal{A}}_\text{even/odd}(t)\,,
\end{equation}
then its time-reversed phasor representation will not be simply even/odd but also complex conjugated:
$$\vc{\mathcal{A}}(-t)=\pm\Re(\vc{A}\ee^{\ii\omega t})=\pm\Re(\vc{A}^*\ee^{-\ii\omega t})\,.$$ The electric dipole is defined as a charge times displacement, both of which should be unchanged by reversing time; hence $\vc{p}\mapsto\vc{p}^*$. The magnetic dipole is a closed current loop multiplied by the area it encloses, and since the current reverses direction when time reverses, we have $\vc{m}\mapsto-\vc{m}^*$. Similarly, $\vc{E}\mapsto\vc{E}^*$ and $\vc{B}\mapsto-\vc{B}^*$ because they are generated by static and moving charges, respectively. Using these simple rules, we can confirm that both $\langle{\vc{F}}_\text{int}\rangle$ and $\langle{\vc{\varGamma}}\rangle=\langle{\vc{\varGamma}_\text{int}}\rangle+\langle{\vc{\varGamma}_\text{rec}}\rangle$ are even under time reversal, exactly as one would expect from forces and torques. Similarly, using the rules for dipoles and incident fields, the extinguished power ($\langle{P_\text{ext}}\rangle$ in \cref{eq:dipolepower}) is time reversal odd $\langle{P_\text{ext}}\rangle\mapsto-\langle{P_\text{ext}}\rangle$. The surprising terms are the scattered power, depending on $\abs{\vc{p}}^2$ and $\abs{\vc{m}}^2$, making it time-reversal-even $\langle{P_\text{sca}}\rangle\mapsto\langle{P_\text{sca}}\rangle$, and the recoil force
$$\langle{\vc{F}}_\text{rec}\rangle=-\tfrac{k^4\eta}{12\pi}\Re(\vc{p}^\ast\!\!\cp\!\vc{m})\mapsto\tfrac{k^4\eta}{12\pi}\Re(\vc{p}\!\cp\!\vc{m}^\ast)=-\langle{\vc{F}}_\text{rec}\rangle\,,
$$
which is odd because it is the product of the electric and magnetic dipoles, which have opposite behaviour under time reversal. As a consequence, the recoil force is \textit{irreversible}, much like frictional or dissipative forces that depend on velocity and convert mechanical energy into heat. This seems particularly surprising, given that the Lorentz force, which can be used to derive this recoil force, is time reversal even:
\begin{equation}
    \label{eq:lorentzforce}
    \vc{F} =  q_\text{e}(\vc{\mathcal{E}} + \vc{v}\times\vc{\mathcal{B}}) + q_\text{m}(\vc{\mathcal{B}}-\tfrac{1}{c^2}\vc{v}\times\vc{\mathcal{E}})\,,
\end{equation}
{even in the generalised case given above with magnetic charges,} since the electric charge $q_\text{e}$ and the electric field $\vc{\mathcal{E}}$ are even, while the magnetic charge $q_\text{m}$, magnetic field $\vc{\mathcal{B}}$, and the velocity $\vc{v}$ are time reversal odd.\footnote{In this work, we assume the magnetic charge to be in $[q_\text{m}]=\si{\ampere\meter}$, such that the magnetic dipole $\vc{m}=q_\text{m}\vc{d}_\text{m}$, which has dimensions of $[\vc{m}]=\si{\ampere\meter\squared}$, is analogous to the electric dipole $\vc{p}=q_\text{e}\vc{d}_\text{e}$.} 

\section{Derivation of recoil terms from the Lorentz force}
{Our aim is to compute the dipolar recoil force from first principles, using the Lorentz force between charges, in order to clearly understand its origin and its time-odd behaviour. This requires using a model of charges and/or currents that implements an electric and magnetic dipole. Doing it in general would be a very cumbersome analytical procedure. Instead, we choose a specific scenario (circularly polarised orthogonal but coplanar electric and magnetic dipoles, realised via co-rotating electric and magnetic point charges) that will greatly simplify the analytical calculation of the Lorentz force, while still allowing us to discern the origin of the recoil force and its time-symmetry. The correctness and validity of the recoil force is not in question, as it has been derived in the past from fundamental momentum conservation arguments, so we do not need to test it generally. Instead we wish to understand its microscopic origin, so considering a specific example is enough for this purpose.} 

Consider a system composed of four charges all moving along a circular trajectory of radius $R$ in the $xy$ plane at the same angular velocity $\omega$.
Each charge $q_i$, with $i\in\{+,\textsc{n},-,\textsc{s}\}$, is placed at a different angle $\alpha_i$ along the circle, such that the position vector of $q_i$ is:
\begin{equation}\label{eq:positionofcharges}
\begin{split}
    \vc{r}_{i}(t) = R \left[ \cos (\alpha_i + \omega t) \uv{x} + \sin (\alpha_i + \omega t) \uv{y} \right].
\end{split}
\end{equation}

\begin{figure}[htb!]
    \centering
    \includegraphics[width=0.55\linewidth]{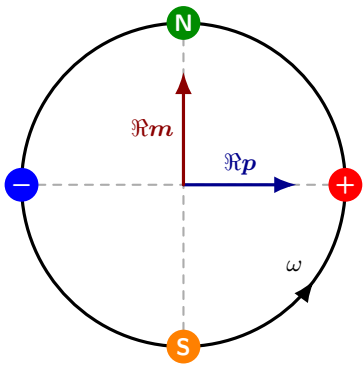}
    \caption{Huygens dipole constructed from electric, $q_\pm=\pm q_\text{e}$, and  magnetic (north/south) charges, $q_\textsc{n/s}=\pm q_\text{m}$, where we assume that $q_\text{e}$, and $q_\text{m}$ are positive. } 
    \label{fig:dipole}
\end{figure}%

A circularly polarised electric dipole $\vc{p} = q_\text{e}\vc{d}_\text{e} = 2R q_\text{e} (\uv{x} + \ii \uv{y})$ is achieved by the two electric charges $q_\pm=\pm q_\text{e}$ at opposite ends of the circular motion ($\alpha_+ = 0$, $\alpha_- = \pi$). At the same time, a circularly polarised magnetic dipole that is in quadrature phase with respect to the electric dipole, that is, $\vc{m} = q_\text{m} \vc{d}_\text{m} = 2 Rq_\text{m} (\uv{y} - \ii \uv{x})=q_\text{m}(-\ii \vc{d}_\text{e})$, can be achieved by adding two \emph{magnetic} point charges $q_\textsc{n/s}=\pm q_\text{m}$ at right angles from the electric charges and also at opposite ends of the circular motion ($\alpha_\textsc{n/s} =\pm \pi/2$). Even though isolated magnetic charges do not exist in nature, we can, for mathematical convenience, regard them as the notional endpoints of a magnetic dipole. This system will, according to \cref{eq:dipoleforce}, have a strong recoil force $\langle{\vc{F}_\text{rec}}\rangle\propto-\Re(\vc{p}^\ast\!\!\cp\!\vc{m})$ acting along the $-z$ direction, particularly if we choose $\abs{\vc{p}} = \abs{\vc{m}}/c$ (or equivalently $q_\text{e}=q_\text{m}/c$), known as Kerker's condition \cite{Liu2018OE, Kerker1982JOSA}. Dipoles satisfying this condition are called Huygens dipoles, and they radiate directionally, in this case in the $z$ direction, with no radiation along $-z$. {The specific system modelled here is therefore a circularly-polarised Huygens dipole, and the arrangement of charges is shown in \cref{fig:dipole}. The directional radiation gives rise to the recoil force.} We aim to calculate this net recoil force by adding up the Lorentz force acting on each of the four charges from the other three, as well as the self-force of each charge on itself \cite{Griffiths2012}.

In order to calculate the electromagnetic fields that a source charge $i$ creates on another charge $j$, we need to take into account the finite speed of light. At any time $t$, the observation point (corresponding to the $j$-th charge) does not see the $i$-th charge where it is now, $\vc{r}_i(t)$, but rather where it was at an earlier retarded time $\vc{r}_i(t - \upDelta t_{ij})$, which predates the current instant by a time interval $\upDelta t_{ij}$. This needs to be calculated self-consistently such that the information, travelling at the speed of light $c$, will reach from the source $\vc{r}_i(t - \upDelta t_{ij})$ to the observer $\vc{r}_j(t)$ in a time $\upDelta t_{ij}$:
\begin{equation}\label{eq:retardedtimecondition}
    c \upDelta t_{ij} = \abs{ \vcrc_{ij}}= \abs{ \vc{r}_{j}(t) - \vc{r}_i(t - \upDelta t_{ij}) }\,.
\end{equation}
{This can be visually interpreted in a space-time diagram (see \cref{fig:lightcone}). 
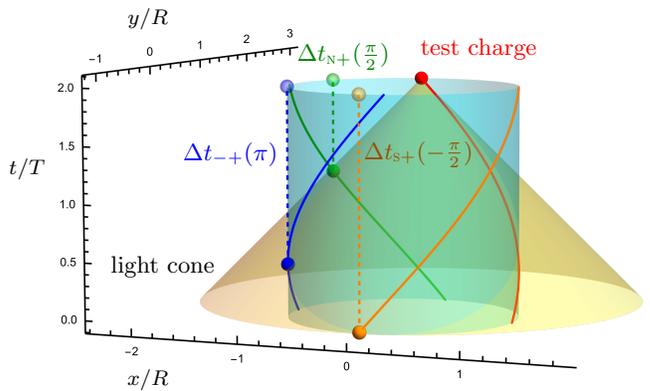
\begin{figure}[t!h]
    \centering
\begin{tikzpicture}
  \node[anchor=south west,inner sep=0] at (0.4,0)
    {\includegraphics[width=.9\linewidth]{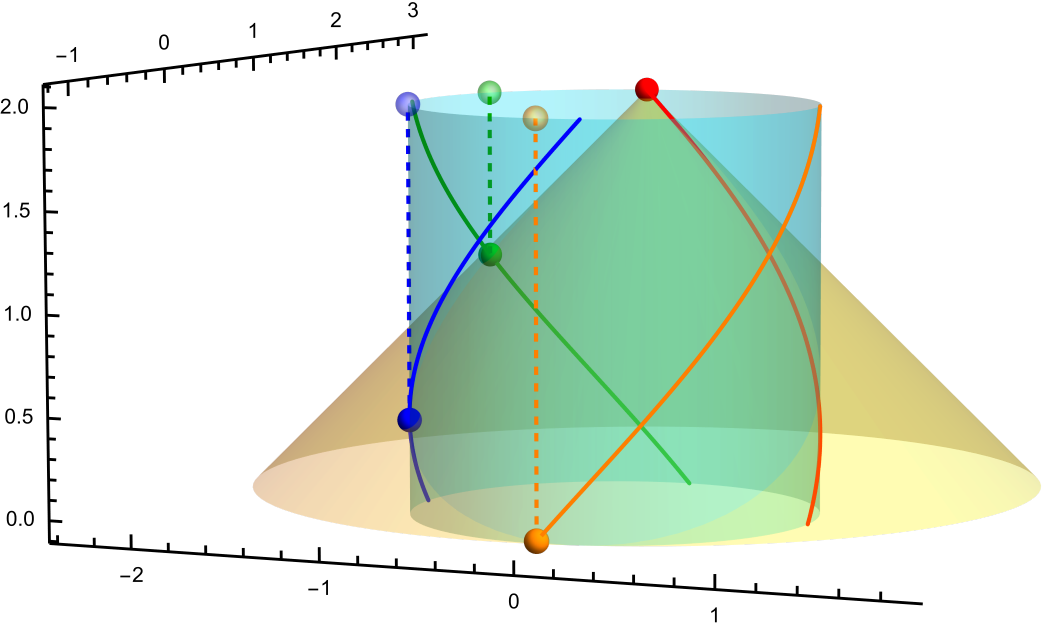}};
  \node at (1.8,1.5) {light cone};
  \node[red] at (6,4.4) {test charge};
  \node[font=\footnotesize] at (1.6,0) {$x/R$};
  \node[font=\footnotesize] at (1.6,4.8) {$y/R$};
  \node[font=\footnotesize] at (0,2.8) {$t/T$};

  \node[orange!60!black] at (5.2,3) {$\upDelta{t}_{\textsc{s}+}(-\tfrac{\pi}{2})$};
  \node[blue] at (2.7,3) {$\upDelta{t}_{-+}(\pi)$};
  \node[green!50!black] at (4.2,4.3) {$\upDelta{t}_{\textsc{n}+}(\tfrac{\pi}{2})$};
\end{tikzpicture}
    \caption{Spacetime diagram of four charges constrained to move on a circular trajectory (shown by the blue cylindrical surface). The solid coloured curves represent the world lines of the charges. The past light cone of the red charge is shown in yellow. Its intersections with the other world lines define the corresponding retarded events (opaque points), whose spatial projections on the circular orbit (retarded positions) are shown as transparent markers. The dashed lines indicate the retarded times along each world line.} 
    \label{fig:lightcone}
\end{figure}%
Each charge, acting as an observer, `sees' the other charges not where they are now, but where they were when the charge's world-line intersected with the observer's past light-cone.} %
\begin{figure}[!tb]
    \centering
    \includegraphics[width=\linewidth]{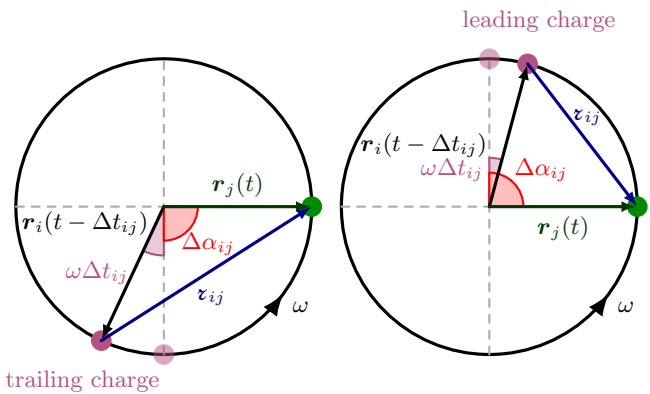}
    \caption{Figure illustrating retarded positions of leading and trailing source charges $\vc{r}_i(t-\upDelta{t}_{ij})$ (denoted by black arrows) as seen from the position of the observer test charge at $\vc{r}_j(t)$ (green arrow), the separation vector $\vcrc_{ij}$ (blue arrow), separation angle $\upDelta\alpha_{ij}$  (red arc), and finally the angle by which the charge position is retarded $\omega\upDelta t_{ij}$ (purple arc).} 
    \label{fig:retardation}
\end{figure}%
Since the retarded position $\vc{r}_i(t - \upDelta t_{ij})$ is on the circumference at an angle given by $\alpha_i + \omega (t-\upDelta t_{ij})$ while the observation point, $\vc{r}_{j}(t)$, is on the same circumference at an angle $\alpha_j + \omega t$, the distance between them is the chord subtended at the origin by an angle $\upDelta\alpha_{ij} - \omega \upDelta t_{ij}$. The length of the chord is given by:
\begin{equation}\label{eq:lengthofchord}
    \abs{ \vcrc_{ij}} = 2R \abs{\sin(\frac{\upDelta\alpha_{ij} - \omega \upDelta t_{ij}}{2})}.
\end{equation}
Equating \cref{eq:retardedtimecondition,eq:lengthofchord} allows solving for $\upDelta t_{ij}$ via reversion of series (see \cref{app:retard}). 
{It is crucial to notice that the retardation will be \emph{different} for the leading and trailing charges (see \cref{fig:lightcone,fig:retardation}), even though their actual instantaneous locations are at equal distances, because the observation charge is moving towards the leading charge, but away from the trailing one.}

Knowing the retarded position, the electric and magnetic fields produced by the electric point charge $q_i$ at the observation charge $q_j$ can be  calculated using the retarded Lienard-Wiechert potentials:
\begin{equation}\label{eq:LWpote}
    \!\!\!\varphi_i(\vc{r}_j,t)=\frac{1}{4\pi\varepsilon_0}\frac{q_i}{(\rc_{ij}-\vcrc_{ij}\vdot\vc{\beta}_{ij})}\,,\quad\!\!\vc{\mathcal{A}}(\vc{r}_j,t)=\frac{\varphi_i\vc{\beta}_{i}}{c}\,,\!\!\!
\end{equation}
where $\vc{v}_{i}=c\vc{\beta}_{i}$ is the relative velocity of the charge $q_i$. All of these are evaluated at the retarded time. The retarded potentials of the magnetic point charge $q_i$ as experienced by a charge $q_j$ can be obtained by applying the duality rotation:
\begin{equation}\label{eq:LWpotm}
    \!\!\psi_i(\vc{r}_j,t)=\frac{\mu_0}{4\pi}\frac{q_i}{(\rc_{ij}-\vcrc_{ij}\vdot\vc{\beta}_{ij})}\,,\quad\!\vc{\mathcal{C}}_i(\vc{r}_j,t)={\psi_i\vc{v}_{ij}}\,.\!\!
\end{equation}
These can then be used to find the fields (for more details see \cref{app:force}):
\begin{equation}\label{eq:retarder_fields}
\begin{split}
    \vc{\mathcal{E}}_i(\vc{r}_j,t)&=-\grad{\varphi}_i-\pdv{\vc{\mathcal{A}}_i}{t}-\curl{\vc{\mathcal{C}}_i}\,,\\
    \vc{\mathcal{B}}_i(\vc{r}_j,t)&=-\grad{\psi}_i-\frac{1}{c^2}\pdv{\vc{\mathcal{C}}_i}{t}+\curl{\vc{\mathcal{A}}_i}\,.
\end{split}
\end{equation}
Knowing the net fields acting on each charge from every other charge, the force acting on it can be computed via the Lorentz force \cref{eq:lorentzforce}. {From electromagnetism's principle of superposition,} the total force in the system will be the sum of sixteen terms, where each charge experiences influence from the retarded positions of every other charge and from itself:
\begin{equation}
    \vc{F} = \sum_i\sum_j\vc{F}_{ij} \quad \text{where} \quad i,j\in\{+,-,\textsc{n},\textsc{s}\}\,.
\end{equation}
Forces on opposite charges will be equal in magnitude and opposite in direction:
\begin{equation}
    \vc{F}_{+-} + \vc{F}_{-+} =\vc{F}_{\textsc{ns}} + \vc{F}_{\textsc{sn}} = 0\,,
\end{equation}
{consistent with the fact that an isolated electric or magnetic dipole does not experience a recoil force on itself.} Similarly, the self-interaction Abraham–Lorentz forces of the opposite charges will be equal and opposite 
\begin{equation}
    \vc{F}_{++} + \vc{F}_{--} = \vc{F}_\textsc{nn} + \vc{F}_\textsc{ss} = 0\,,
\end{equation}
since they are proportional to the square of the charge and its acceleration, and it is these accelerations that will be opposite. That leaves eight mixed terms arising from interactions between electric and magnetic charges. {These remaining terms do not cancel each other (e.g. $\vc{F}_\textsc{+n} + \vc{F}_\textsc{n+} \neq 0$) purely due to the fact that leading and trailing charges are retarded by different amounts. The \cref{app:force} details all algebraic steps leading to the final net force on the system of charges, precisely matching the expression for the recoil force}:
\begin{equation}
    \begin{split}
        \vc{F}&=-\frac{k^4\eta_0}{12\pi} (q_\text{e}q_\text{m}8R^2)\uv{z}=-\frac{k^4 \eta_0}{12\pi}\Re(\vc{p}^\ast\!\!\cp\!\vc{m})\,.
    \end{split}
\end{equation}

From this calculation, it is also easier to see why the force is time reversal odd and what this means. We can see that time reversal will not only make the whole system spin in the opposite sense and exchange the places of north and south charges (this would already explain the time reversal odd behaviour), but it will also turn the retarded potentials \cref{eq:LWpote,eq:LWpotm} into advanced potentials. This means that the force no longer depends on the past position of the other charges; in fact, it depends on the future position. Of course, this will make sense only if the observer is watching this system in reverse. Let us look at the power \cref{eq:dipolepower}, since we assume no illumination $\langle{P_\text{ext}}\rangle=0$, which means the system will not be absorbing power but losing it to the radiation $\langle{P_\text{abs}}\rangle=-\langle{P_\text{sca}}\rangle$. The force $\langle{\vc{F}_\text{rec}}\rangle$ is the recoil from this radiation, ensuring momentum conservation. Looking also at the torque $\expval{\vc{\Gamma}_\text{rec}}$, we can see that it acts against the initial rotational motion of the dipoles. Therefore, the Huygens dipole will convert its rotational motion into radiation and translational motion until it loses all its energy. In the time-reversed picture, the dipole will be absorbing power $\langle{P_\text{abs}}\rangle=\langle{P_\text{sca}}\rangle$ (see \cref{sec:symmetries}), which comes from the scattered fields that will be returning to the dipole in this perspective. The force $-\langle{\vc{F}_\text{rec}}\rangle$ is now the force coming from the dipole being pushed by the scattered fields absorbed by the system. Since the $\expval{\vc{\Gamma}_\text{rec}}$ stays the same, but the sense of rotation of the dipole changes, the rotation will be speeding up. This, of course, makes sense if we think of time-reversal as playing the video of this phenomenon backwards.

\section{Conclusions}
In this work, we have rederived the recoil optical force acting on an electromagnetic dipole directly from the Lorentz force between the constituent moving charges, rather than relying on the Maxwell stress tensor. This first-principles approach reproduces the established recoil terms while clarifying their physical origin and symmetry properties under time reversal.

We demonstrated that the recoil force naturally arises from the retarded interactions between the electric and magnetic charges forming the dipoles, with the asymmetry between leading and trailing charges being essential to generate a net non-zero Lorentz force. This retardation breaks time-reversal symmetry, making the recoil term odd under time reversal, in contrast with the interaction terms, which are even. As a result, the recoil force behaves analogously to dissipative forces, reflecting the irreversible conversion of mechanical energy into radiated electromagnetic momentum.

Furthermore, the comparison of time-reversal symmetries across power, force, and torque reveals a consistent physical picture. While the force and power contain both even and odd terms under time reversal, the torque’s interaction and recoil components are both even. In the physical sense, this corresponds to the recoil torque opposing the system’s angular momentum in the forward-time process, while in the time-reversed picture the torque acts in the same direction as the rotation, causing it to accelerate. This symmetry relationship is consistent with intuition about how the system’s rotational and translational motions evolve under time reversal.

Altogether, our derivation connects the microscopic Lorentz-force picture with the macroscopic recoil phenomena, providing clear physical insight into how an isolated Huygens dipole exchanges momentum and angular momentum with its own radiation. These results consolidate the understanding of recoil forces as self-induced yet momentum-conserving effects, relevant to self-propelling and radiation-driven dipolar systems.
\begin{acknowledgments}
SG and FJRF are supported by the EIC-Pathfinder-CHIRALFORCE (101046961) which is funded by Innovate UK Horizon Europe Guarantee (UKRI project 10045438). NL would like to thank Pietro Polenghi for useful discussions.   
\end{acknowledgments}

\hbadness 10000\relax\bibliography{main}

\onecolumngrid

\section*{Supplementary Information}
\appendix
\crefalias{section}{appendix}    
\crefalias{subsection}{appendix}

\section{Retarded time for two arbitrary charges moving around a circle at a constant angular frequency}\label{app:retard}
Consider a test charge and a source charge, both moving at angular velocity $\omega$ around a circle of radius $R$. Without loss of generality, we can describe the position of the test charge as:
\begin{equation}\label{eq:obs}
    \vc{r}_j(t)= R [ \cos(\omega t) \uv{x} + \sin(\omega t) \uv{y}]\,,
\end{equation}
and the position of the source charge as:
\begin{equation}
   \vc{r}_i(t)=R [ \cos(\omega {t}+\upDelta\alpha_{ij}) \uv{x} + \sin(\omega {t}+\upDelta\alpha_{ij}) \uv{y} ]\,.
\end{equation}
Note that the angle $\upDelta\alpha_{ij}$ can be negative as the source charge can either lead the test charge ($0<\upDelta\alpha_{ij}<\pi$), be exactly opposite ($\upDelta\alpha_{ij}=\pi$) or trail it ($-\pi<\upDelta\alpha_{ij}<0$). In the case studied later (see \cref{fig:retardation}), $\upDelta\alpha_{ij}$ will be $\pi/2$ (leading source charge), $\pi$ (diametrically opposite charge), or $-\pi/2$ (trailing source charge); however, here $\upDelta\alpha_{ij}$ is kept arbitrary.

Taking into account the finite speed of light the fields acting on the test charge at the observation position $ \vc{r}_j(t)$ need to be evaluated as produced from the source charge located in the retarded position at the retarded time $t-\upDelta{t}_{ij}$:
\begin{equation}\label{eq:ret_position}
   \vc{r}_i(t-\upDelta{t}_{ij})=R [ \cos(\omega {t}-\omega \upDelta{t}_{ij}+\upDelta\alpha_{ij}) \uv{x} + \sin(\omega {t}-\omega \upDelta{t}_{ij}+\upDelta\alpha_{ij}) \uv{y} ]\,,
\end{equation}
where $\upDelta{t}_{ij}$ is the time it takes for light to travel between the source $\vc{r}_i(t - \upDelta t_{ij})$ and the observation point $\vc{r}_j(t)$:
\begin{equation}
    c \upDelta{t}_{ij} = \abs{ \vc{r}_j(t) - \vc{r}_i(t - \upDelta t_{ij}) }\,.\label{eq:retardedtimecondition2}
\end{equation}
The appearance of $\upDelta t_{ij}$ on both sides of this equation means that it must be solved self-consistently. Taking \cref{eq:obs,eq:ret_position} and the sum and difference trigonometric formulas, the separation vector is:
\begin{equation*}
   \vcrc_{ij}(t)=\vc{r}_j(t)-\vc{r}_i(t-\upDelta{t}_{ij})=2R\sin(\frac{\upDelta\alpha_{ij}-\omega \upDelta{t}_{ij}}{2}) \left[ \sin(\omega {t}+\frac{\upDelta\alpha_{ij}-\omega \upDelta{t}_{ij}}{2}) \uv{x} - \cos(\omega {t}+\frac{\upDelta\alpha_{ij}-\omega \upDelta{t}_{ij}}{2}) \uv{y} \right],
\end{equation*}
this means that \cref{eq:retardedtimecondition2} can be rewritten for phase retardation as:
\begin{equation}\label{eq:retardationequation}
    \omega\upDelta{t}_{ij}= 2\beta\abs{\sin(\frac{\upDelta\alpha_{ij}-\omega \upDelta{t}_{ij}}{2})} \,,
\end{equation}
where we used the fact that the orbital velocity $v=\beta c=\omega R$.
We will want to take the dipole limit (where the size of the system is much smaller than the wavelength); this is essentially the limit when the orbital speed is non-relativistic:
\begin{equation}
    kR=\frac{2\pi R}{\lambda}=\frac{\omega R}{c}\ll1\quad \Leftrightarrow \quad\beta\ll1\,.
\end{equation}
Since the retarded phase vanishes when $\beta=0$ we can expect it to have the following Taylor expansion:
\begin{equation}\label{eq:omega delta t function alpha}
    \omega\upDelta{t}_{ij}=a_1\beta+a_2\beta^2+a_3\beta^3+a_4\beta^4+\order{\beta^5}\,,
\end{equation}
where we expand to the fourth order since \cref{eq:recforcephasor} is of order $k^4=(\beta/R)^4$.
We can plug this expansion into \cref{eq:retardationequation}, expand the sine function, and then solve order by order for coefficients $a_n$:
\begin{align*}
    a_1&=2\abs{\sin(\tfrac{1}{2}{\upDelta\alpha_{ij}})}\,,\\
    a_2&=-\sin(\upDelta\alpha_{ij})\,,\\
    a_3&=\tfrac{1}{2}\qty[1+3\cos({\upDelta\alpha_{ij}})]\abs{\sin(\tfrac{1}{2}{\upDelta\alpha_{ij}})}\,,\\
    a_4&=\tfrac{1}{3}\qty[\sin(\upDelta\alpha_{ij})-2\sin(2\upDelta\alpha_{ij})]\,.
\end{align*}
The retarded phase for the opposite charge $\upDelta\alpha_{ij}=\pi$ is:
\begin{equation}
    \omega\upDelta{t}_{ij}(\pi)=2\beta-\beta^3+\order{\beta^5}\,.
\end{equation}
This reproduces the expected result that, at the lowest order, we have precisely the distance between the charges $c\upDelta{t}_{ij}(\pi)\approx2R$.
For neighbouring charges separated by $\upDelta\alpha_{ij}=\pm\pi/2$, we have:
\begin{equation}\label{eq:retardedneighbours}
    \omega\upDelta{t}_{ij}\qty(\pm\frac{\pi}{2})=\sqrt{2}\beta\mp\beta^2+\frac{\beta^3}{2\sqrt{2}}\pm\frac{\beta^4}{3}+\order{\beta^5}\,,
\end{equation}
which is again at the lowest order $c\upDelta{t}_{ij}(\pm{\pi}/{2})\approx R\sqrt{2}$ matching the expected result. A crucial point is that the even orders have opposite signs for the leading and trailing charges, therefore the two will have different retarded times $\upDelta{t}_{ij}\qty({\pi}/{2})\neq\upDelta{t}_{ij}\qty(-{\pi}/{2})$, which means that the force on the leading and trailing neighbouring charges will be different. This difference is ultimately responsible for a net recoil force in the system, as we will show in the next section.

\section{Lorentz force of a Huygens dipole from first principles}
\label{app:force}
This supplementary will derive the Lorentz force of a circularly-polarised Huygens dipole from first principles using the retarded fields \cref{eq:retarder_fields}. The fields produced by an electric charge $q_i$ and experienced by a charge $q_j$ are as follows:
\begin{equation}
        \vc{\mathcal{E}}_{\text{e},i}(\vc{r}_j,t) = \frac{q_i}{4 \pi \epsilon _0}\frac{|\vcrc_{ij}| }{({\vcrc_{ij}  \vdot \vc{u}_{ij}})^3}[\underbrace{(c^2 - v^2)\vc{u}_{ij}}_\text{near field} + \underbrace{{\vcrc}_{ij}\times(\vc{u}_{ij}\times\vc{a}_{i})}_\text{far field}]\Big|_{\mathrm{ret}}\,,\quad\vc{\mathcal{B}}_{\text{e},i}(\vc{r}_j,t) = \frac{1}{c}[\uvrc_{ij}\cp\vc{\mathcal{E}}_{\text{e},i}(\vc{r}_j,t)]\Big|_{\mathrm{ret}}\,,\label{eq:original elec}
\end{equation}
and similarly, a magnetic charge $q_i$ will produce the fields:
\begin{equation}
        \vc{\mathcal{B}}_{\text{m},i}(\vc{r}_j,t) = \frac{\mu_0q_i}{4\pi}\frac{|\vcrc_{ij}| }{({\vcrc_{ij} \vdot \vc{u}_{ij}})^3}[\underbrace{(c^2 - v^2)\vc{u}_{ij}}_\text{near field} + \underbrace{{\vcrc}_{ij}\times(\vc{u}_{ij}\times\vc{a}_{i})}_\text{far field}]\Big|_{\mathrm{ret}} \,,\quad\vc{\mathcal{E}}_{\text{m},i}(\vc{r},t) = -c[\uvrc_{ij}\cp\vc{\mathcal{B}}_{\text{m},i}(\vc{r}_j,t)]\Big|_{\mathrm{ret}}\,,\label{eq:original mag}
\end{equation}
where the retarded separation vector $\vcrc_{ij}$ is given by
\begin{equation}
    \vcrc_{ij}(t)=\vc{r}_j(t)-\vc{r}_i(t-\upDelta{t}_{ij}),\label{eq: retarded separation vector definition}
\end{equation}
We recall $\vc{r}_j(t)$ is the position vector of the test charge at the observation point, and $\vc{r}_i(t-\upDelta{t}_{ij})$ is the position of the retarded source charge that produces a field at point $\vc{r}_j(t)$. We define the vector $\vc{u}_{ij}$ as
\begin{equation}
    \vc{u}_{ij}(t) = c {\uvrc}_{ij}(t) - \vc{v}_{i}(t-\upDelta{t}_{ij}), \label{eq: retarded velocity vector definition}
\end{equation}
where the retarded separation unit vector is $\hat\vcrc_{ij} = \vcrc_{ij}/|\vcrc_{ij}|$, the retarded source velocity is $\vc{v}_{i} = {\dv{t}}\vc{r}_{i}$, and the retarded acceleration is $\vc{a}_{i} ={\dv{t}}\vc{v}_{i}$. We will also use the generalised Lorentz force equation to determine the force experienced by each charge 
\begin{equation}
    \vc{F}_{ij}(\vc{r}_j,t) =  q_{\text{e},j}[\vc{\mathcal{E}}_i(\vc{r}_j,t) + \vc{v}_j\times\vc{\mathcal{B}}_i(\vc{r}_j,t)] + q_{\text{m},j}[\vc{\mathcal{B}}_i(\vc{r}_j,t) -\varepsilon_0\mu_0\vc{v}_j\times\vc{\mathcal{E}}_i(\vc{r}_j,t)],\label{eq:force 1}
\end{equation}
where $\vc{v}_j$ in this case is the velocity of a charge at the observation point $\vc{r}_j(t)$, and we must consider the total electric field $\vc{\mathcal{E}}_i=\vc{\mathcal{E}}_{\text{e},i} +\vc{\mathcal{E}}_{\text{m},i}$ and total magnetic field $\vc{\mathcal{B}}_i = \vc{\mathcal{B}}_{\text{e},i} + \vc{\mathcal{B}}_{\text{m},i}$. However, since our system has no particles with both electric and magnetic charge, both $\vc{\mathcal{E}}_i$ and $\vc{\mathcal{B}}_i$ will only have one of the two (e/m) contributions.

\subsection{Determining the General Force Expression}\label{sec: general force}
Since each of the four charges in the Huygens dipole exert forces on each of the other charges and themselves, the total force of the Huygens dipole will consist of a sum of sixteen terms denoted by the following:
\begin{equation}
    \vc{F} = \sum_i\sum_j\vc{F}_{ij} \quad \text{where} \quad i,j\in\{+,-,\textsc{n},\textsc{s}\}\,.
\end{equation}
Note that by symmetry, it can be determined that the electric-electric and magnetic-magnetic forces will be equal and opposite and hence sum to zero:
\begin{equation}
    \vc{F}_{+-} + \vc{F}_{-+} =\vc{F}_{\textsc{ns}} + \vc{F}_{\textsc{sn}} = 0.
\end{equation}
Likewise, we can demonstrate that the sum of the four Abraham-Lorentz self-forces of each of the charges of the Huygens dipole in \cref{fig:dipole} will equal to zero. First, note that the Abraham-Lorentz formula reads:
\begin{equation}
    \vc{F}^\text{e}_\text{self} = \frac{q_\text{e}^2}{6\pi \varepsilon_0 c^3}\dot{\vc{a}},\qq{or} \vc{F}^\text{m}_\text{self} = \frac{\mu_{0}q_\text{m}^2}{6\pi c^3}\dot{\vc{a}}
\end{equation}
where $\dot{\vc{a}}$ is the first time derivative of the acceleration of the charge in question. At any instant, it can be shown that the $\dot{\vc{a}}$ vectors for the electric charges ($\dot{\vc{a}}_{++}=-\dot{\vc{a}}_{--}$) and the $\dot{\vc{a}}$ vectors of the magnetic charges ($\dot{\vc{a}}_\textsc{nn}=-\dot{\vc{a}}_\textsc{ss}$) are equal and opposite, therefore, the four Abraham-Lorentz self-forces of the Huygens dipole will sum to zero:
    \begin{equation}
        \vc{F}_{++} + \vc{F}_{--} = \vc{F}_\textsc{nn} + \vc{F}_\textsc{ss} = 0.
    \end{equation}
Thus, we are left with only the four electric-magnetic and the four magnetic-electric force contributions with a non-zero sum.

This would not be the case if the effects of retardation were not accounted for since the distance between the leading and trailing charges to a reference charge would be equal (see \cref{fig:retardation}). Due to this, and the fact that leading and trailing charges have opposite signs, the force on the reference charge due to the leading charge would be equal and opposite to that of the trailing charge and hence the total sum of the forces of the rotating Huygens dipole would be zero. If retardation is taken into account however, then the retarded distance between the leading and trailing charges to the reference charge is no longer equal (see \cref{fig:lightcone,fig:retardation}), hence, the forces due to them experienced by the reference charge are no longer equal and opposite, and therefore the sum of these forces is non-zero. Following the method laid out above, and using Wolfram Mathematica to handle the algebraic steps in \cref{sec: general fields,sec: near and far field}, we find the final force expression to be:
\begin{equation}
    \vc{F} = -\frac{q_eq_m}{4\pi \epsilon_0 c}\frac{8\beta^4}{3R^2}\uv{z}.
\end{equation}
This net force originates from and acts on the system itself, thus violating the law of conservation of mechanical momentum and therefore Newton's third law. Total momentum is conserved however as this arrangement of charges radiates away electromagnetic momentum that exactly accounts for the loss of mechanical momentum of the system.

\subsection{Electric-to-magnetic and Magnetic-to-electric contributions}\label{sec: general fields}
The force contributions that do not cancel are the following:
\begin{equation}
    \vc{F} = \underbrace{\overbrace{\vc{F}_{+\textsc{n}}+\vc{F}_{-\textsc{s}}}^\text{trailing}+\overbrace{\vc{F}_{+\textsc{s}}+\vc{F}_{-\textsc{n}}}^\text{leading}}_\text{Electric-to-magnetic}+\underbrace{\overbrace{\vc{F}_{\textsc{n}+}+\vc{F}_{\textsc{s}-}}^\text{leading}+\overbrace{\vc{F}_{\textsc{s}+}+\vc{F}_{\textsc{n}-}}^\text{trailing}}_\text{Magnetic-to-electric}
\end{equation}
and they can be separated into groups depending on which type of charge (electric or magnetic) is the source and depending on whether the souce charge is leading or trailing. One can use the second half of \cref{eq:original elec} and \cref{eq:force 1} to show that the {electric-to-magnetic} contributions (electric source charge $q_i$ and magnetic target charge $q_j$) will be of the form:
\begin{equation}\label{eq:Fplus}
    \vc{F}{\kern0.05em}^\text{e}_{i j} = \frac{q_j}{c^2}(c^2\vc{\mathcal{B}}_{i j}-\vc{v}_j\times\vc{\mathcal{E}}_{i j})=\frac{q_j}{c^2}(c\uvrc_{i j}-\vc{v}_j)\times\vc{\mathcal{E}}_{i j}\,,
\end{equation}
where $\vc{\mathcal{E}}_{i j}=\vc{\mathcal{E}}_{i}(\vc{r}_j)$ and $\vc{\mathcal{B}}_{i j}=\vc{\mathcal{B}}_{i}(\vc{r}_j)$. Similarly, \cref{eq:original mag} and \cref{eq:force 1} lead to magnetic-to-electric components (magnetic source $q_i$ and electric target $q_j$) having a very similar expression:
\begin{equation}\label{eq:Fminus}
    \vc{F}{\kern0.05em}^\text{m}_{ij} = {q_j}(\vc{\mathcal{E}}_{ij}+\vc{v}_j\times\vc{\mathcal{B}}_{ij})=-{q_j}(c\uvrc_{ij}-\vc{v}_j)\times\vc{\mathcal{B}}_{ij}\,.
\end{equation}
Notice that we have added a superscript to the force to indicate whether it is electric-to-magnetic ($\vc{F}{\kern0.05em}^\text{e}_{i j}$) or magnetic-to-electric component ($\vc{F}{\kern0.05em}^\text{m}_{i j}$).
The reason we have done that is that if we directly substitute for $\vc{\mathcal{E}}_{i j}$ in \cref{eq:Fplus} from \cref{eq:original elec} and likewise $\vc{\mathcal{B}}_{i j}$ in \cref{eq:Fminus} from \cref{eq:original mag} we obtain a general expression:
\begin{equation}\label{eq:Fplusminus}
        \vc{F}{\kern0.05em}^\text{e,m}_{ij} = \sigma_\text{e,m}\frac{\mu_0{q}_i{q}_j}{4 \pi }\frac{(c\uvrc_{ij}-\vc{v}_j)\times[{(c^2 - v^2)\vc{u}_{ij}} + {{\vcrc}_{ij}\times(\vc{u}_{ij}\times\vc{a}_{i})}]}{|\vcrc_{ij}|^2({\uvrc_{ij}  \vdot \vc{u}_{ij}})^3}\,,
\end{equation}
where $\sigma_\text{e}=1$ and $\sigma_\text{m}=-1$ is a sign that depends on whether the source $q_i$ is an electric or magnetic charge, respectively. At this stage, the best we can do is explicitly evaluate \cref{eq:Fplusminus} for a pair of charges. As a reminder from \cref{app:retard}, we can describe the position of any test charge as:
\begin{equation*}
    \vc{r}_j(t)= R [ \cos(\omega t) \uv{x} + \sin(\omega t) \uv{y}]\,,
\end{equation*}
and the retarded position of source charge as
\begin{equation*}
   \vc{r}_i(t-\upDelta{t}_{ij})=R [ \cos(\omega {t}-\omega \upDelta{t}_{ij}+\upDelta\alpha_{ij}) \uv{x} + \sin(\omega {t}-\omega \upDelta{t}_{ij}+\upDelta\alpha_{ij}) \uv{y} ]\,,
\end{equation*}
their separation will be
\begin{equation*}
   \vcrc_{ij}(t)=\vc{r}_j(t)-\vc{r}_i(t-\upDelta{t}_{ij})=2R\sin(\frac{\upDelta\alpha_{ij}-\omega \upDelta{t}_{ij}}{2}) \left[ \sin(\omega {t}+\frac{\upDelta\alpha_{ij}-\omega \upDelta{t}_{ij}}{2}) \uv{x} - \cos(\omega {t}+\frac{\upDelta\alpha_{ij}-\omega \upDelta{t}_{ij}}{2}) \uv{y} \right],
\end{equation*}
and the remaining vectors are defined in terms of these:
\begin{equation*}
   \vc{v}_i=\dv{\vc{r}_i}{t},\quad\vc{v}_j={\dv{\vc{r}_j}{t}},\quad\vc{a}_i={\dv{\vc{v}_i}{t}},\quad\vc{u}_{ij} = c {\uvrc}_{ij} - \vc{v}_{i},\quad \uvrc_{ij} = \frac{\vcrc_{ij}}{|\vcrc_{ij}|}\,.
\end{equation*}
which can all be substituted into \cref{eq:Fplusminus}. After lengthy algebra handled by Wolfram Mathematica, we arrive at the expression:
\begin{equation}\label{eq:Findeltaalpha}
        \vc{F}{\kern0.05em}^\text{e,m}_{ij} = \sigma_\text{e,m}\frac{{q}_i{q}_j}{4 \pi c\varepsilon_0 R^2}\uv{z}\underbrace{\qty[{\sin(\upDelta\alpha_{ij}-\omega\upDelta{t}_{ij})+\frac{2}{\beta}\abs{\sin(\frac{\upDelta\alpha_{ij}-\omega\upDelta{t}_{ij}}{2})}}]^{-1}}_{\displaystyle f_{ij}(\upDelta\alpha_{ij})}.
\end{equation}
\Cref{eq:Findeltaalpha} together with the $\beta$ expansion for $\upDelta{t}_{ij}(\upDelta\alpha_{ij})$ in \cref{eq:omega delta t function alpha} can now be used to obtain every non-zero term of the force. At this point, we can appreciate the fact that each of these terms is independent of $t$ and depends only on $\upDelta\alpha_{ij}-\omega\upDelta{t}_{ij}$, so we can equate the instantaneous force on the system $\vc{F}$ with its time average $\langle\vc{F}\rangle$. In all of these terms the charges are separated by $\upDelta\alpha_{ij}=\pm\pi/2$ leading to the expansion:
\begin{equation}
    f_{ij}\qty(\pm\frac{\pi}{2})=\frac{\beta}{\sqrt{2}}-\frac{\beta^3}{4\sqrt{2}}\pm\frac{\beta^4}{3}+\order{\beta^5}\,,
\end{equation}
Finally, we can obtain the force due to trailing charges, which are all equal to:
\begin{equation}
    \vc{F}_{+\textsc{n}}=\vc{F}_{-\textsc{s}}=\vc{F}_{\textsc{s}+}=\vc{F}_{\textsc{n}-}= \frac{{q}_\text{e}{q}_\text{m}}{4 \pi \varepsilon_0c R^2}f_{ij}\qty(-\frac{\pi}{2})\uv{z}\,,
\end{equation}
and similarly leading charges, which are equal:
\begin{equation}
    \vc{F}_{\textsc{n}+}=\vc{F}_{\textsc{s}-}=\vc{F}_{+\textsc{s}}=\vc{F}_{-\textsc{n}}=- \frac{{q}_\text{e}{q}_\text{m}}{4 \pi \varepsilon_0c R^2}f_{ij}\qty(\frac{\pi}{2})\uv{z}\,,
\end{equation}
and indeed they are not equal, nor opposite as $f_{ij}\qty(-\frac{\pi}{2})\neq \pm f_{ij}\qty(\frac{\pi}{2})$. Adding all of these terms together leads to the final expression for the total force:
\begin{equation}\label{eq:rectot}
    \vc{F}= -\frac{{q}_\text{e}{q}_\text{m}}{\pi \varepsilon_0c R^2}\qty[f_{ij}\qty(\frac{\pi}{2})-f_{ij}\qty(-\frac{\pi}{2})]\uv{z}=-\frac{{q}_\text{e}{q}_\text{m}}{\pi \varepsilon_0c R^2}\qty[\frac{2\beta^4}{3}+\order{\beta^5}]\uv{z}\,,
\end{equation}
which in the dipolar limit ($R\ll\lambda$, hence $\beta=v/c=2\pi R/\lambda\ll1$) agrees with the well-known \cref{eq:recforcephasor} obtained for this setup from the usual expression derived from Maxwell's stress tensor.


\subsection{Determining the Near and Far Field Contributions}\label{sec: near and far field}
We can repeat the calculation for the near and far field components of \cref{eq:original elec,eq:original mag} separately:
\begin{equation}
    \vc{F}=\frac{{q}_\text{e}{q}_\text{m}}{\pi \varepsilon_0c R^2}\Big[\underbrace{\qty(\beta^2-\frac{19\beta^4}{6})}_\text{near field}\underbrace{\mathop{-}\qty(\beta^2-\frac{5\beta^4}{2})}_\text{far field}+\order{\beta^5}\Big]\uv{z}\,,
\end{equation}
doing so we can see that while they partly cancel out, they don't cancel out completely, and both contribute to the total force.

\section{Dipole phasors}
\subsection{Force and torque}
The relationship between a time harmonic physical quantity $\vc{\mathcal{A}}(t)$ and its phasor representation $\vc{A}$ is 
$$\vc{\mathcal{A}}(t)=\Re(\vc{A}\ee^{-\ii\omega t})\,.$$ 
One of the many advantages of this representation is that it allows finding time-averaged quantities (quadratic in fields) such as power, force, torque, etc. The time average is performed over one period $T=2\pi/\omega$, for example if $\vc{F}(t)$ is an instantaneous force, then $\expval{\vc{F}}=\frac{1}{T}\int_0^T\vc{F}(t)\dd{t}$. One can show that if you have two time-harmonic fields $\vc{\mathcal{A}}$ and $\vc{\mathcal{B}}$ with phasor representations $\vc{A}$ and $\vc{B}$ we have:
\begin{equation}
    \expval{\vc{\mathcal{A}}\cp\vc{\mathcal{B}}}=\frac{1}{T}\int_0^T\vc{\mathcal{A}}(t)\cp\vc{\mathcal{B}}(t)\dd{t}=\frac{1}{2}\Re(\vc{A}^*\cp\vc{B})\,.
\end{equation}
Using the phasor representation of dipoles, the recoil force, torque, and scattered power are
\begin{align}
    \expval{\vc{F}_\text{rec}}&=-\frac{k^4}{12\pi\varepsilon_0c}\Re(\vc{p}^*\cp\vc{m})\,,\label{eq:force}\\
    \expval{\vc{\Gamma}_\text{rec}}&=-\frac{k^3}{12\pi }\Im\qty(\frac{1}{\varepsilon_0}\vc{p}^*\!\cp\vc{p}+\mu_0\vc{m}^*\!\cp\vc{m})\label{eq:torque},\\
    \expval{P_\text{sca}}&=\frac{\omega k^3}{12\pi}\qty(\frac{1}{\varepsilon_0}\abs{\vc{p}}^2\!+\!\mu_0\abs{\vc{m}}^2).
\end{align}
\subsection{Spinning Huygens dipole}
The electric and magnetic dipoles placed at right angle with each other, spinning in the anticlockwise sense, can be represented by the following phasors:
\begin{equation}
    \vc{p}=q_\text{e}(2R)({\uv{x}+\ii\uv{y}})=q_\text{e}\vc{d}\,,\quad
    \vc{m}=q_\text{m}(2R)({\uv{y}-\ii\uv{x}})=-\ii q_\text{m}\vc{d}\,,
\end{equation}
where $d=2R$ is the separation of the two opposite charges and $q_\text{e}$, $q_\text{m}$ are respectively the electric and magnetic charges. It is worth noticing that the phasor $(\uv{x}+\ii\uv{y})$ represents a vector field that is spinning anticlockwise:
\begin{equation}
    \Re[(\uv{x}+\ii\uv{y})\ee^{-\ii\omega t}]=(\uv{x}\cos{\omega t}+\uv{y}\sin{\omega t})\,.
\end{equation}
The cross product of $\vc{d}=2R({\uv{x}+\ii\uv{y}})$ and its conjugate is:
\begin{equation}
    {\vc{d}^*\cp\vc{d}}=4R^2(\uv{x}-\ii\uv{y})\cp(\uv{x}+\ii\uv{y})=4R^2[{\ii\uv{x}\cp\uv{y}-\ii\uv{y}\cp\uv{x}}]=8\ii R^2 \uv{z}.
\end{equation}
Now all that is left to do is plug these dipoles into \cref{eq:force,eq:torque}. Doing this for the recoil force we find it indeed matches \cref{eq:rectot}:
\begin{equation}\label{eq:recforcephasor}
    \begin{split}
        \expval{\vc{F}_\text{rec}}&=-\frac{k^4}{12\pi\varepsilon_0c}\Re[q_\text{e}(-\ii q_\text{m})\vc{d}^*\cp\vc{d}]=-\frac{2}{3\pi\varepsilon_0c} q_\text{e}q_\text{m}(k^4R^2)\uv{z}=-\frac{2}{3\pi\varepsilon_0c}q_\text{e}q_\text{m}\qty(\frac{\beta^4}{R^2})  \uv{z}\,.
    \end{split}
\end{equation}
where in the last line we used the definition of $\beta=v/c=\omega R/c=kR$. The same can be done for the torque:
\begin{equation}
    \begin{split}
        \expval{\vc{\Gamma}_\text{rec}}=-\frac{2}{3\pi}\qty(\frac{q_\text{e}^2}{\varepsilon_0}+{\mu_0\,q_\text{m}^2})(k^3R^2)\uv{z}=-\frac{2}{{3\pi}}\qty(\frac{q_\text{e}^2}{\varepsilon_0}+{\mu_0\, q_\text{m}^2} )\qty(\frac{\beta^3}{R}) \uv{z}\,.
    \end{split}
\end{equation}
The calculation of dipolar torque using Lorentz forces, for an isolated electric and magnetic dipole, is a textbook problem that agrees with the result above (Ref. \cite{Griffiths2012}). To generalise this to a combination of electric and magnetic dipole, we must simply add the torque contributions $\vc{\varGamma}_{ij}^{\text{e,m}}=\vc{r}_j\times\vc{F}{\kern0.05em}^\text{e,m}_{ij}$ caused by the electric-to-magnetic and magnetic-to-electric forces $\vc{F}{\kern0.05em}^\text{e,m}_{ij}$ that were computed exactly above, and can be shown to cancel out pairwise, hence not contributing to the net torque. Finally, the power scattered by the spinning Huygens dipole:
\begin{equation}
    \begin{split}
         \expval{P_\text{sca}}=\frac{2c}{3\pi}\qty(\frac{q_\text{e}^2}{\varepsilon_0}+\mu_0q_\text{m}^2)k^4=\frac{2c}{3\pi}\qty(\frac{q_\text{e}^2}{\varepsilon_0}+\mu_0q_\text{m}^2)\qty(\frac{\beta}{R})^4.
    \end{split}
\end{equation}

\end{document}